\documentclass[prb,twocolumn,showpacs,floatfix]{revtex4}

\usepackage{graphicx}

\begin{document}
 \title{Two-Point Entanglement Near a Quantum Phase Transition}
\author{Han-Dong Chen}
\affiliation{Department of Physics, University of Illinois at Urbana-Champaign, Urbana, IL 61801}
 \begin{abstract}
In this work, we study  the two-point entanglement $S(i,j)$, which measures the entanglement between two separated degrees of freedom $(ij)$ and the rest of system, near a quantum phase transition. Away from the critical point,  $S(i,j)$ saturates with a characteristic length scale $\xi_E$, as the distance $|i-j|$ increases. The entanglement length $\xi_E$ agrees with the correlation length.  The universality and finite size scaling of entanglement are demonstrated in a class of exactly solvable one dimensional spin model. By connecting the two-point entanglement to correlation functions in the long range limit, we argue that the prediction power of a two-point entanglement is universal as long as the two involved points are separated far enough.
 \end{abstract}
\pacs{03.67.Mn, 75.30.Kz, 89.75.Da}
 \maketitle

Quantum phase transition happens at zero temperature when physical parameters are changed\cite{sachdevbook}. 
Just like the classical critical phenomena\cite{cardybook}, a diverging length scale dominates the physics near a quantum phase transition. Quantum criticality also exhibits scaling law and universality class. 
Meanwhile, quantum system contains a new physics, entanglement\cite{Nielsonbook}. Entanglement is a unique property of quantum system and measures quantum correlations. It is thus natural to ask if entanglement possesses the dominating correlations near a quantum phase transition. Tremendous interests and work have been invested along this line. In particular, entanglement of formation between two separate spins after tracing out the rest has been studied widely\cite{Arnesen2001,OConnor2001,Osterloh2002,Osborne2002,Zanardi2002,Dusuel2004}. It is shown that this quantity shows pronounced behaviors such as scaling law and universality class\cite{Osterloh2002,Dusuel2004}. However, this quantity vanishes as two spins are separated by several lattice constants. Thus, it does not contain the dominating long-range feature between two separated spins possessed by the system. The entanglement between a block of spins and the rest of system is also studied in various models\cite{Vidal2003,Refael2004,Calabrese2004,Calabrese2005,Latorre2005,Unanyan2005}.    However, there is no direct relation between it and the long range correlation of the system either\cite{Verstraete2004}. Recently, Verstraete et al\cite{Verstraete2004} studied the average entanglement that can be localized between two separated spins by performing local measurements on the other individual spins.  The typical length scale at which the localizable entanglement decays is defined as the entanglement length.
All classical correlation functions provide lower bounds for localizable entanglement. However, the entanglement length defined for localizable entanglement can be infinite even for a gapped system\cite{Verstraete2004a}. Therefore, an entanglement measure describing the correct long range physics near a continuous phase transition has yet to be studied. It is the purpose of this work to identify and study such a measure of entanglement near a quantum phase transition.

We start by noticing that two-point correlation functions can be essential in the study of quantum phase transitions. It is also convenient to organize correlation functions into a reduced density matrix. We will thus focus on the  possible entanglements encoded in a reduced density matrix. The reduced density matrix of two separated points $i$ and $j$ contains at least three types of entanglement.  The first type is the entanglement between $i$ (or $j$) and the rest of system. It can be measured by the Von Neumann entropy of the reduced density matrix of single point $i$, which in turn can be obtained from the two-point reduced density matrix by partial tracing out $j$. This quantity has been studied near quantum phase transition\cite{Osborne2002} and does not contain an explicit length scale. The second type is the entanglement between  $i$ and $j$ and can be measured by concurrence for spin-1/2 systems. It shows beautiful scaling law  but does not contain a diverging entanglement length scale near a quantum phase transition\cite{Osterloh2002}. In this work, we shall focus on the behavior of the third type of entanglement, the entanglement between $(ij)$ and the rest of system.  At zero temperature, the whole system is described by a pure state. It is thus justified to use the Von Neumann entropy of the two-point reduced density matrix to measure the entanglement between the two points $(ij)$ and the rest of system. We shall call this type of entanglement as two-point entanglement. It is shown that two-point entanglement is dominated by a diverging length scale in the vicinity of a quantum critical point. 
We show the entanglement length diverges as approaching a quantum phase transition and agrees with the correlation length in an exactly solvable spin model. We also demonstrate that two-point entanglement exhibits beautiful scaling law and universality\cite{Osborne2002,Cao2006}.  The non-analyticity of the two-point entanglement for two nearest-neighboring points  has been successfully utilized to efficiently identify possible phase transitions \cite{Wu2004a,Zanardi2002a,Gu2004,Legeza2006,Cao2006}. Because of the nice scaling properties, phase transition point can be determined from small systems with considerable accuracy without pre-assumed order parameters\cite{Cao2006}. However, studies along this line are so far limited to specific models and it is not clear how universal its prediction power of a possible phase transition is. 
By connecting the two-point entanglement to correlation functions, we shall argue that the prediction power of a two-point entanglement is universal as long as the two involved points are separated far enough. The tow-point entanglement has great advantages in predicting possible phase transitions. Although it does not provide extra information beyond correlation functions, two-point entanglement  automatically picks out the most dominant non-analyticity of all possible correlation functions. Therefore, one single quantity is enough to distinguish phase transitions between different orders. Pre-assumed order parameters are thus not needed. This is particularly convenient when studying systems with complicated phase diagram, for instance, strongly correlated systems such as Hubbard model\cite{Gu2004}.  We also want to point out that two-point entanglement is different from the entanglement entropy of block spins studied widely in literature. The reduced density matrix of two spins at $i$ and $j$ contains only a subset of the information encoded in the reduced density matrix of a block of spins with size of the block $L\geq|i-j|$. Because of the non-additive nature of the entanglement entropy, the behavior of two-pint entanglement studied in this work can be very different from the one of block spins. The entanglement entropy of a block of spins contains more information than our two-point entanglement does. However, it is one of our main results that the two-point entanglement bears enough information to efficiently identify possible quantum phase transitions. The two-point entanglement is also closely related to the mutual information recently studied by Gu et al\cite{Gu2006}. In a translationally invariant system, two-point entanglemtn and mutual information share the same range-dependence and thus bare the same characteristic length scale. However, two-point entanglement has an advantage that it is a better measurement of quantum information in the sense that the mutual information shared by two systems may be larger than the entropy in either part. 

As an example, let us study the two-point entanglement in an spin-1/2 ferromagnetic chain with an exchange $J$ and a transverse field $h$,
\begin{eqnarray}
H=-\sum_{i=1}^N\left[\frac{J(1+\gamma)}{2}\sigma^x_i\sigma^x_{i+1}+
\frac{J(1-\gamma)}{2}\sigma^y_i\sigma^y_{i+1}+h\sigma^z_i\right]
%-h\sum_{i=1}^N ,\nonumber\\&&
\label{Eq-H}
\end{eqnarray}
where $\sigma^x,\sigma^y,\sigma^z$ are Pauli matrices, $N$ is the number of sites and $\gamma$ is the anisotropy in the $xy$ plane. This model belongs to the Ising universality class when $0<\gamma\leq 1$ and $N=\infty$. It can be solved exactly and the correlation functions are given in Ref.\cite{Lieb1961,McCoy1968,Pfeuty1970,Barouch1971}.

The two-point entanglement $S(i,j)$ is defined as the entropy entanglement between two separate spins, $i$ and $j$, and the remaining spins in the system. From the full density matrix of the whole system, we can obtain the reduced density matrix $\rho$ for $i$ and $j$ by partial tracing out the rest spins. Based on the symmetry of the Hamiltonian, we know that $\rho$ is real and symmetric\cite{Osterloh2002}. The nonzero elements are $\rho_{11}, \rho_{22}=\rho_{33},\rho_{44},\rho_{14}$ and $\rho_{23}$ if we choose the basis $|1\rangle=|\uparrow\uparrow\rangle, |2\rangle=|\uparrow\downarrow\rangle,|3\rangle=|\downarrow\uparrow\rangle, |4\rangle=|\downarrow\downarrow\rangle$. These elements can be related to the correlation functions $p^\alpha(i,j)=\langle\sigma^\alpha_i\sigma^\alpha_{j}\rangle-\langle\sigma^\alpha_i\rangle\langle\sigma^\alpha_{j}\rangle$ for $\alpha=x,y,z$ and magnetization $M(i)=\langle \sigma^z_i\rangle$\cite{Osborne2002}.  We can then define the two-point entanglement as the entropy of $\rho$
\begin{eqnarray}
S(i,j)=-Tr \rho \log \rho.
\end{eqnarray}
$S(i,j)$ measures the entanglement between two separated spins, $i$ and $j$, and the rest spins. Since the system is translationally invariant, we know $p^\alpha$ and $S$ only dependent on the distance $n=|i-j|$ and $M$ is site-independent.

\begin{figure}[t]
\includegraphics[width=2.5in]{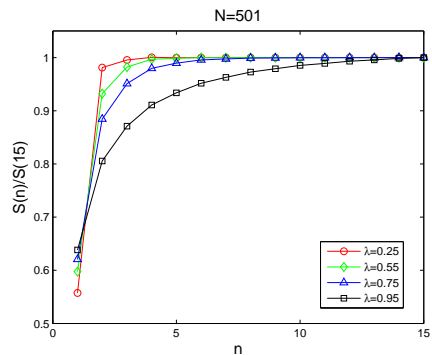}
\caption{S(n) as a function of distance $n$ for different values of $\lambda$ in the Ising limit where $\gamma=1$. As $\lambda$ starts from small value and approaches the critical value $1$, it takes longer and longer distance for $S(n)$ to saturate. A length scale, entanglement length $\xi_E$, can be defined to measure how fast the two-point entropy saturates.}\label{FIG-sn}
\end{figure}

Using the results of Ref.\cite{Lieb1961,McCoy1968,Pfeuty1970,Barouch1971}, we can calculate the reduced density matrix $\rho$ and thus $S$ exactly. Let us introduce the dimensionless tuning parameter $\lambda=J/h$. There is a quantum phase transition at $\lambda_c=1$ for an infinite spin chain. For $\lambda>1$, the system is ferromagnetically ordered and kinks start to condense at $\lambda=1$ which drives the system into a quantum paramagnetic state in $\lambda<1$ region. 

In Fig.\ref{FIG-sn}, we show $S(n)$ as a function of distance $n$ for different values of $\lambda$ in the Ising limit where $\gamma=1$. $S(n)$ increases as $n$ increases. This is opposite to the decaying of correlation functions. We want to emphasize that $S$ measures the entanglement between two spins $(ij)$ and the rest spins while correlation function measures the correlation between $i$ and $j$. Crudely speaking, $S$ can be taken as an accumulation of the correlation attached to $i$ or/and $j$. Consequently, the two-point entanglement $S(n)$ is expected to increase and reach a constant as $n$ increases. Fig.\ref{FIG-sn} reveals a more important message. As $\lambda$ starts from a small value and approaches the critical value $\lambda_c=1$, it takes longer and longer distance for $S(n)$ to reach the long-range limit $S(\infty)$. One can thus define a length scale which measures how fast the two-point entanglement $S$ saturates. This can be shown by examining the long range behavior of $S(n)$ for $\lambda = \lambda_c(N)-\delta\lambda$ near $\lambda_c(N)$. In the thermodynamic limit $N\rightarrow \infty$, the long range behavior of correlation functions has been studied in Ref.\cite{McCoy1968,Pfeuty1970,Barouch1971}.  From Ref.\cite{Barouch1971}, the asymptotic forms of correlation functions for $\lambda<1$ and large $n$ are
\begin{eqnarray}
p^x(n)&\sim& n^{-1/2}\lambda_2^{n}\\
p^y(n)&\sim& n^{-3/2}\lambda_2^{n}\\
p^z(n)&\sim& n^{-2}\lambda_2^{2n}
\end{eqnarray}
with 
\begin{eqnarray}
\lambda_2=\frac{1/\lambda-\sqrt{1/\lambda^2-(1-\gamma^2)}}{1-\gamma}.
\end{eqnarray}
In the limit of $\gamma\rightarrow 1$, we have $\lambda_2\rightarrow\lambda$. Near the critical point $\lambda_c=1$, $\lambda_2\approx 1+(\lambda-1)/\gamma$ to the first order of $\lambda-1$. Since $p^y$ and $p^z$ vanish faster than $p^x$ as $n$ goes to infinity, it is safe to set them to zero for large enough $n$. The eigenvalues of the reduced density matrix $\rho$ are thus
\begin{eqnarray}
\epsilon_{1,2}&=&\left(1-M^2\pm p^x(n)\right)/4 \\
\epsilon_{3,4}&=&\left(1+M^2\pm \sqrt{4M^2+[p^x(n)]^2}\right)/4.
\end{eqnarray}
It follows that
\begin{eqnarray}
S(n)- S(\infty) \propto [p^x(n)]^2\propto n^{-1} e^{-n/\xi_E}\label{Eq-entanglement}
\end{eqnarray}
where the entanglement length $\xi_E$ is given by
\begin{eqnarray}
\xi_E=\frac{\gamma/2}{1-\lambda}.\label{Eq-LE}
\end{eqnarray} 
Eq.(\ref{Eq-entanglement}) directly relates the entanglement to the correlation function. The expression Eq.(\ref{Eq-entanglement}) is essentially a Taylor expansion of $S(n)$ in terms of the correlation functions and thus holds as long as the correlation functions are small, {\it i.e.}, when $n$ is longer than the correlation length. In Fig.\ref{FIG-S-n} (a), we compare the exact results of $S(n)$ and the scaling relation given by Eq.({\ref{Eq-entanglement}) for $\lambda =0.9$. The plot of $\ln (S(n)-S(\infty))$ vs $n$ for $\lambda=0.9$ confirms the entanglement length given by Eq.(\ref{Eq-LE}) directly. It is interesting to notice that $\xi_E$ is directly related to the correlation length. The critical exponent $\nu=1$ is also independent of the anisotropy $\gamma$, in agreement with the universality hypothesis. Similarly, from the asymptotic forms for $\lambda>1$ given in Fig.1 of Ref.\cite{Barouch1971}, we again obtain $\xi_E\propto (\lambda-1)^{-1}$ with critical exponent $\nu=1$. 
%For $\lambda<1$ and $n$ large, $p^x$ reads\cite{Pfeuty1970}
%\begin{eqnarray}
%p^x(n) \approx (1-\lambda^2)^{-1/4} \pi^{-1/2} n^{-1/2}\lambda^n.
%\end{eqnarray}
%$p^y$ and $p^z$ vanish much quicker than $p^x$ near $\lambda_c=1$ when $n$ goes to infinity. We can thus safely set $p^y$ and $p^z$ to zero for simplicity. 
%The eigenvalues of $\rho$ are then given by
%\begin{eqnarray}
% \epsilon_{1,2}&=& \frac{1}{4}\left[1-M^2\pm\frac{\lambda^n}{(1-\lambda^2)^{1/4}\sqrt{n\pi}}\right]\\
% \epsilon_{3,4}&=& \frac{1}{4}\left[1+M^2\pm\sqrt{4M^2 +\frac{\lambda^{2n}}{(1-\lambda^2)^{1/2}n\pi}}\right]
%\end{eqnarray}
%It follows that 
%\begin{eqnarray}
%S(n) \approx S(\infty,\lambda) -  C(\lambda)n^{-1} e^{-n/\xi_E}
%\end{eqnarray}
%where the entanglement length $\xi_E$ is given by
%\begin{eqnarray}
%\xi_E = \frac{1}{2\delta\lambda}=\frac{1}{2(1-\lambda)}\label{Eq-LE}
%\end{eqnarray}
%and $C(\lambda)$ is a non-negative $\lambda$-dependent constant. It is interesting to notice that $\xi_E$ has the same critical exponent $\nu=1$ as the correlation length does. 
Eq.(\ref{Eq-LE}) suggests that the entanglement length diverges at the critical point $\lambda=1$ and the two-point entanglement should approach $S(\infty)$ through a power law. We shall now demonstrate that this is indeed the case. For $\lambda=1$ and large $n$, we have\cite{Barouch1971}
\begin{eqnarray}
p^x(n)&\propto& n^{-1/4}\\
p^y(n)&\propto& n^{-9/4}\\
p^z(n)&\propto& n^{-2}.
\end{eqnarray}
The two-point entanglement can be shown to follow
\begin{eqnarray}
S(n)- S(\infty)\propto  n^{-1/2},\label{Eq-powerlaw}
\end{eqnarray}
where the exponent $-1/2$ is related to the anomalous dimension of $\sigma^x$. 
In Fig.\ref{FIG-S-n} (b), we compare the exact results of $S(n)$ and the scaling relation given by Eq.({\ref{Eq-entanglement}) for $\lambda = 1$. The plot of $\ln (S(n)-S(\infty))$ vs $\ln n$ for $\lambda=0.9$ confirms the exponent $-1/2$ in the scaling law Eq.(\ref{Eq-powerlaw}) directly.

\begin{figure}
\includegraphics[width=2.5in]{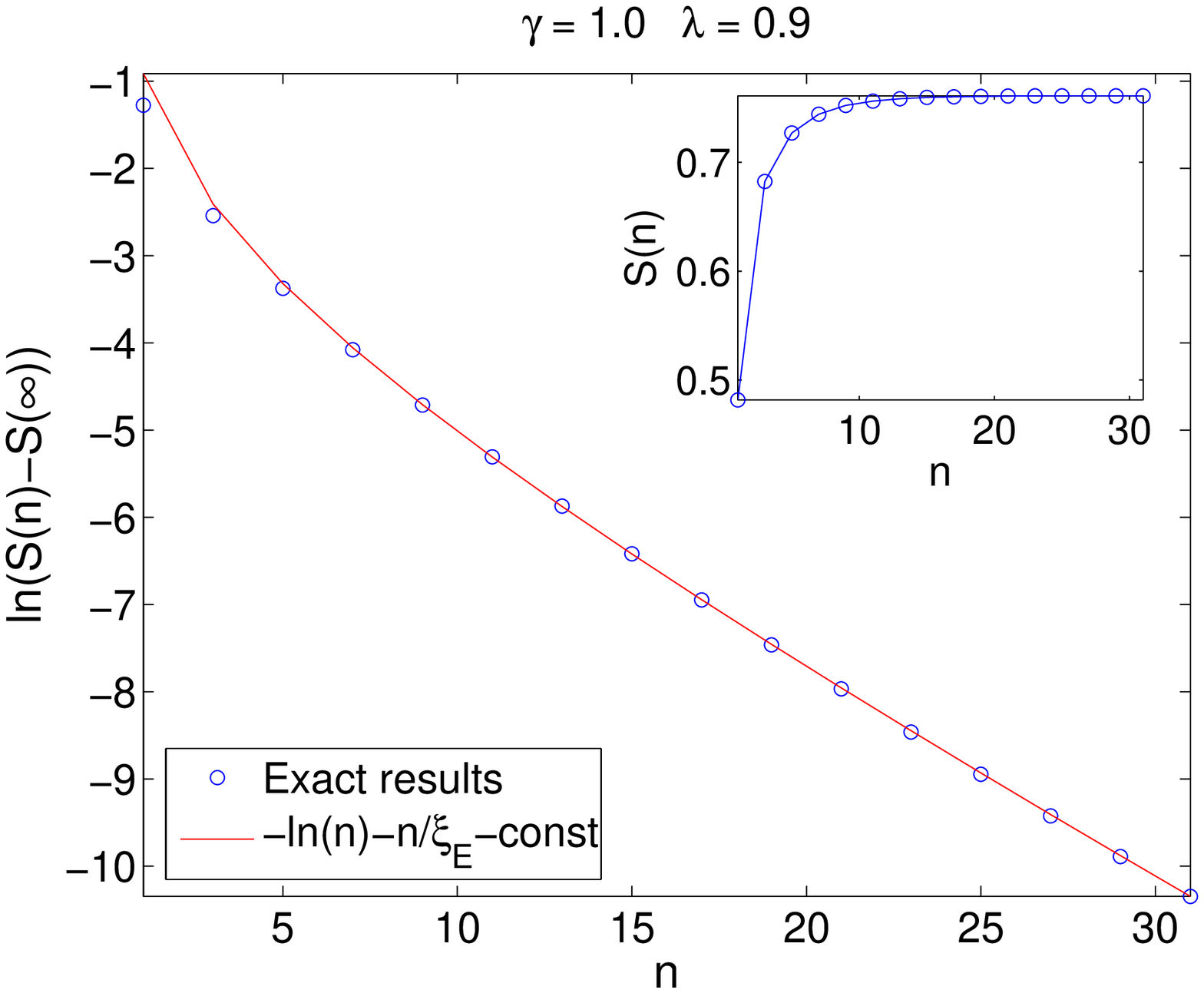}\\ ~ (a) \\ ~ \\
\includegraphics[width=2.5in]{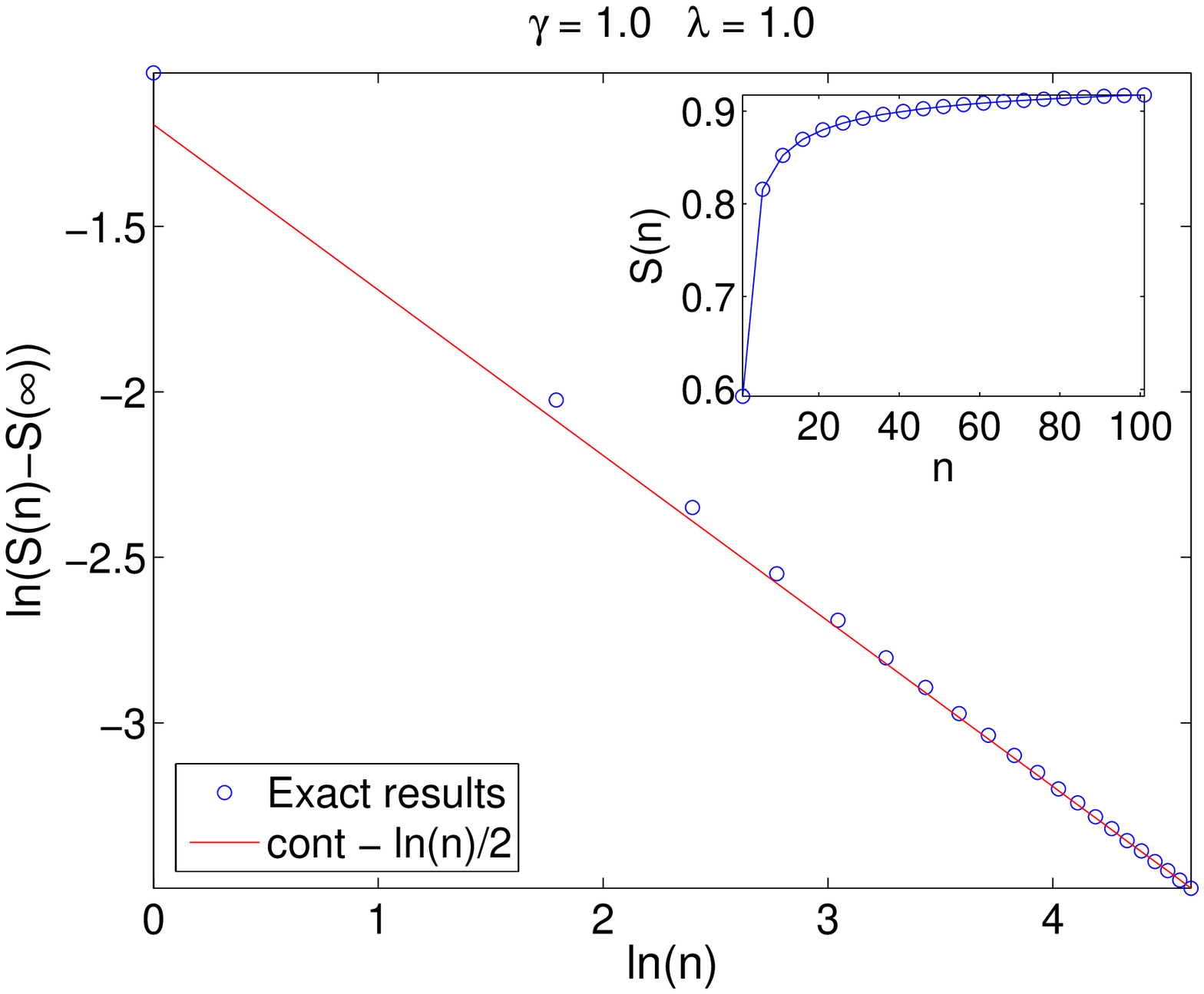}\\ ~ (b) 
\caption{(a) The comparison between exact results of $S(n)$ for $\lambda=0.9$ and Eq.(\ref{Eq-entanglement}) and Eq.(\ref{Eq-LE}).  The $\ln (S(n)-S(\infty))$ vs $n$ plot directly confirms the entanglement length defined in Eq.(\ref{Eq-LE}). Inset is the plot of $S(n)$ as a function of $n$. (b) The comparison of $S(n)$ for $\lambda=1$ and Eq.(\ref{Eq-powerlaw}). The plot of $\ln (S(n)-S(\infty))$ vs $\ln n$ for $\lambda=0.9$ confirms the exponent $-1/2$ in the scaling law Eq.(\ref{Eq-powerlaw}) directly. Inset is the plot of $S(n)$ vs $n$.}\label{FIG-S-n}
\end{figure}

\begin{figure}[t]
\includegraphics[width=2.5in]{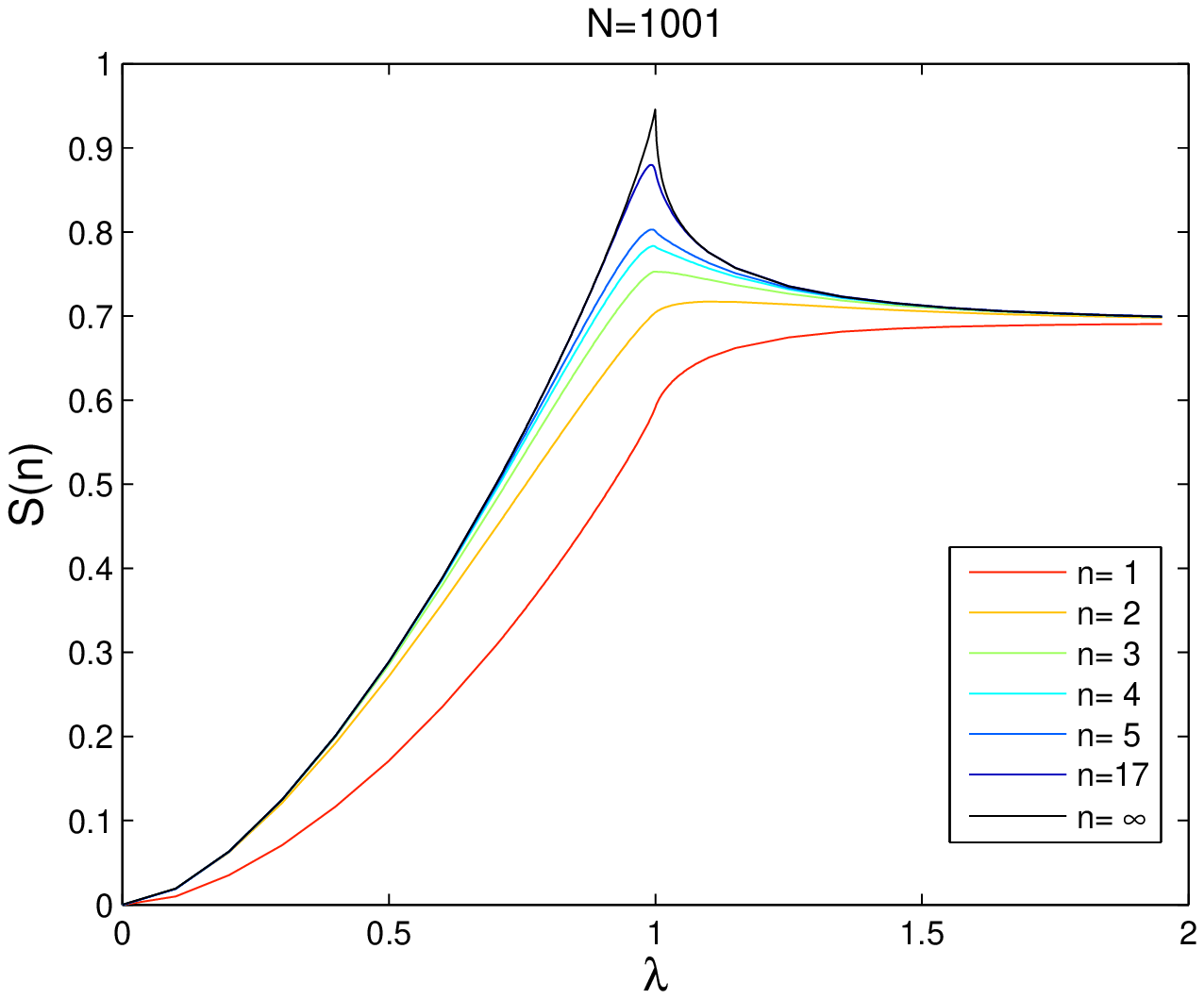} \\
\includegraphics[width=2.5in]{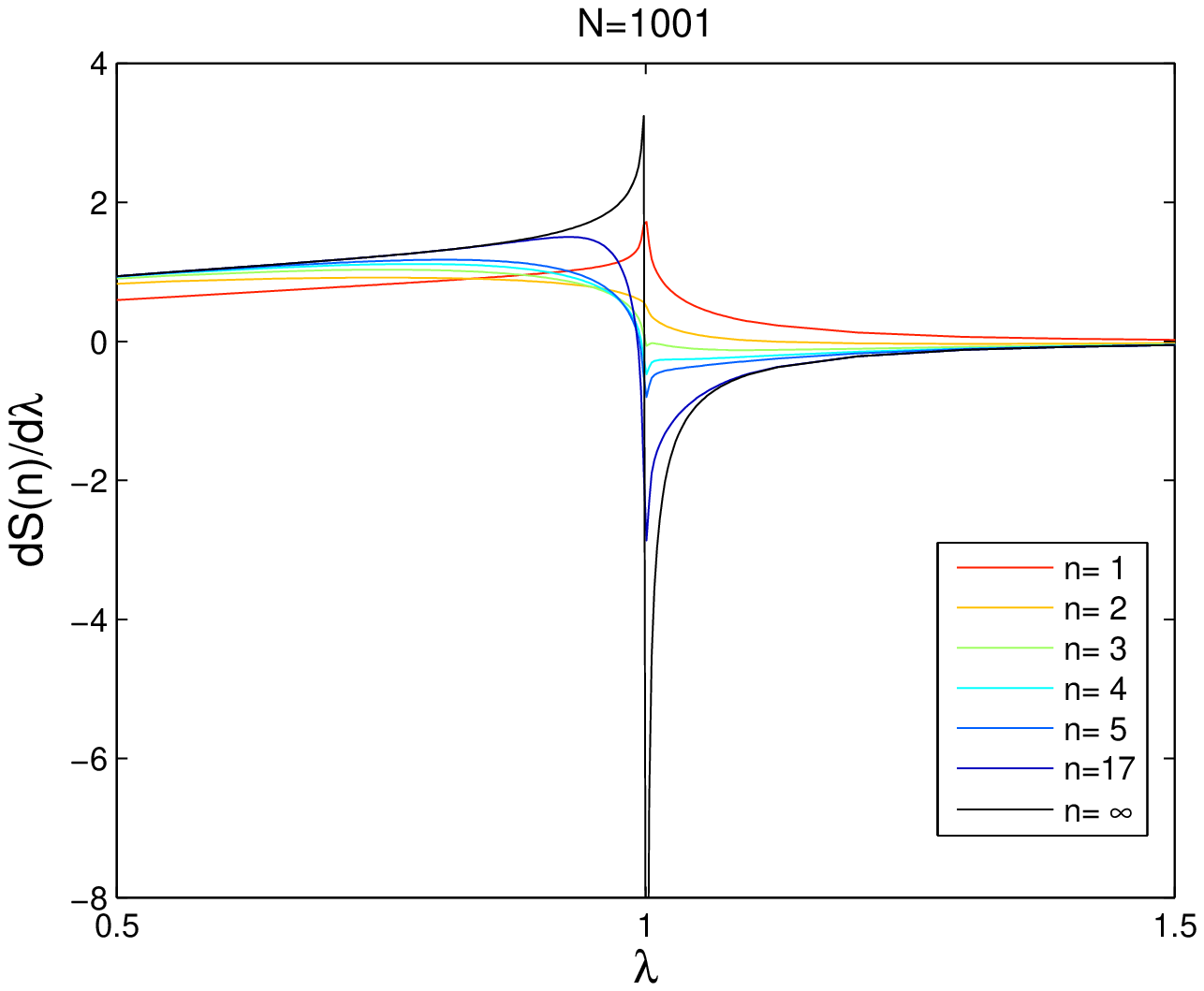}
\caption{Top: $S(n)$; bottom: $dS(n)/d\lambda$ as functions of $\lambda$ for different $n$. For finite $n$, the maximum of $S$ does not correspond to a phase transition while sharp dip or peak in $dS/d\lambda$ does. When $n\rightarrow\infty$ the maximum of $S$ shifts to the critical point and displays a $\lambda$-like anomaly. The anisotropy $\gamma$ is set to $1$ in this plot.}\label{FIG-s-ds}
\end{figure}

We thus arrive at the main result of this work. The two-point entanglement $S(n)$ contains a length scale, the entanglement length $\xi_E$ that agrees with the correlation length.  For $0<\gamma\leq 1$,  $S(n)$ reaches $S(\infty)$ exponentially over the length scale $\xi_E$. $\xi_E$ diverges as $\lambda$ approaches the critical value $\lambda_c=1$ with an exponent $\nu=1$. At $\lambda=1$, $S(n)$ follows a power law given by Eq.(\ref{Eq-powerlaw}) with an exponent $-1/2$. 
Therefore, the two-point entanglement bears the same dominating long range physics about the phase transition as correlation functions do. This is not surprising since the reduced density matrix $\rho$ and consequently the two-point entanglement $S(n)$ are smooth functions of correlation functions. We also notice that the entanglement length $\xi_E$ defined in this work is closely related to the one defined in Ref.\cite{Verstraete2004}, where it is defined as the measure of how fast the localizable entanglement between two separated spins decays as the distance increases. In a very crude model, as we mentioned previously, the two-point entanglement $S(i,j)$ can be taken as an accumulation of entanglement. Only those spins that are entangled with one or both of $i$ and $j$ can contribute. Therefore, through the study of the distance dependence of the two-point entanglement, we learn how the entanglement is shared out among spins. If the localizable entanglement attached to either spin decays exponentially with a length scale $\xi$, one then expects $S(n)$ increases as $n$ increases but still shorter than $\xi$. After $n>>\xi$, $S(n)$ stays at a constant as further separation no longer introduces extra localizable entanglement. However, the contribution of localizable entanglement to the two-point entanglement is in general not greater than the optimized one defined in Ref.\cite{Verstraete2004}. Thus, the two definitions of entanglement length can be very different in some cases. For instance, it has been shown that the decay length of localizable entanglement can be infinite in a gapped system\cite{Verstraete2004a}. In contrast, the entanglement length defined in this work is expected to follow the correlation length which is finite\cite{Fan2004}.

\begin{figure}
\includegraphics[width=2.5in]{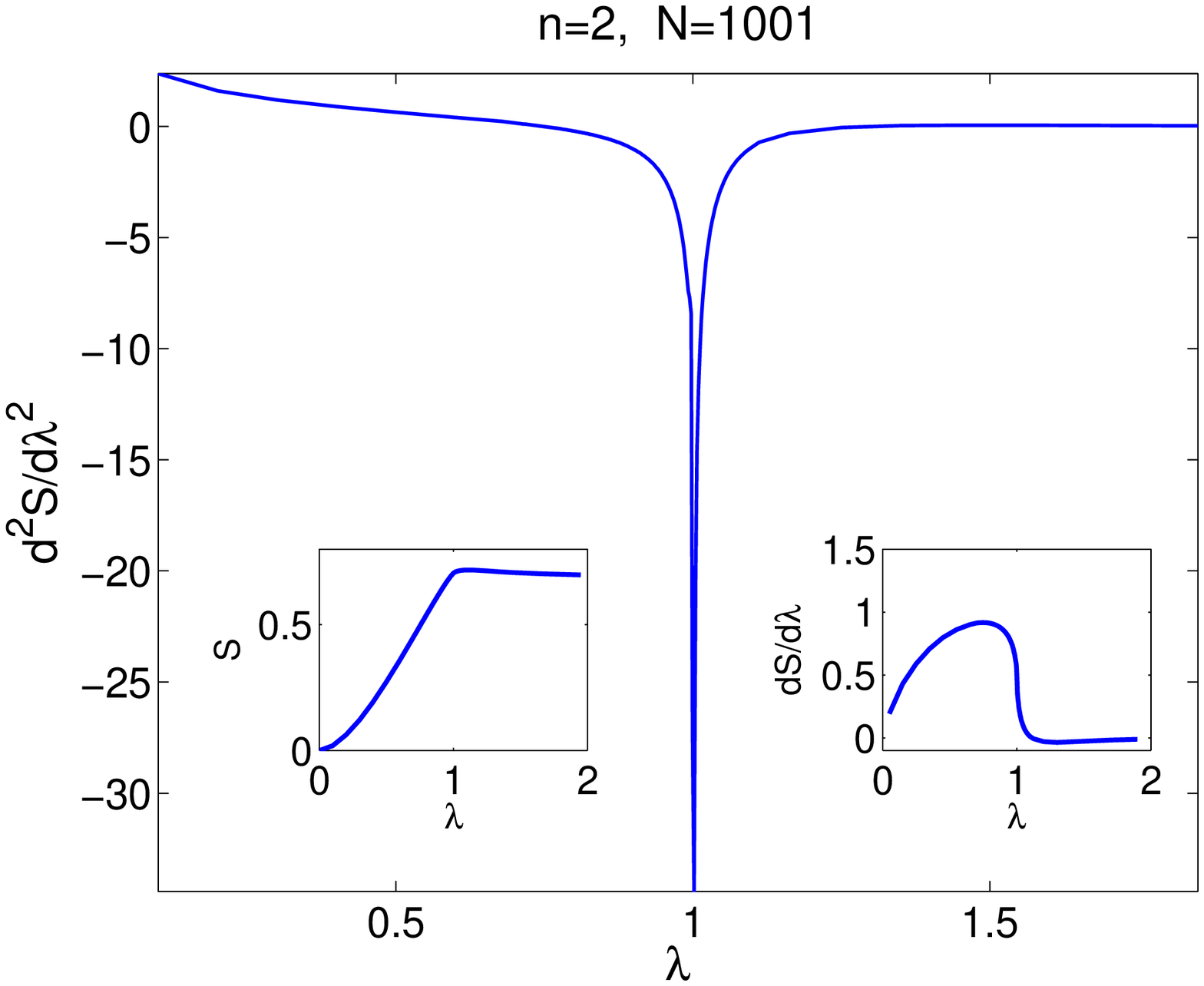} \\ (a) \\ ~ \\
\includegraphics[width=2.5in]{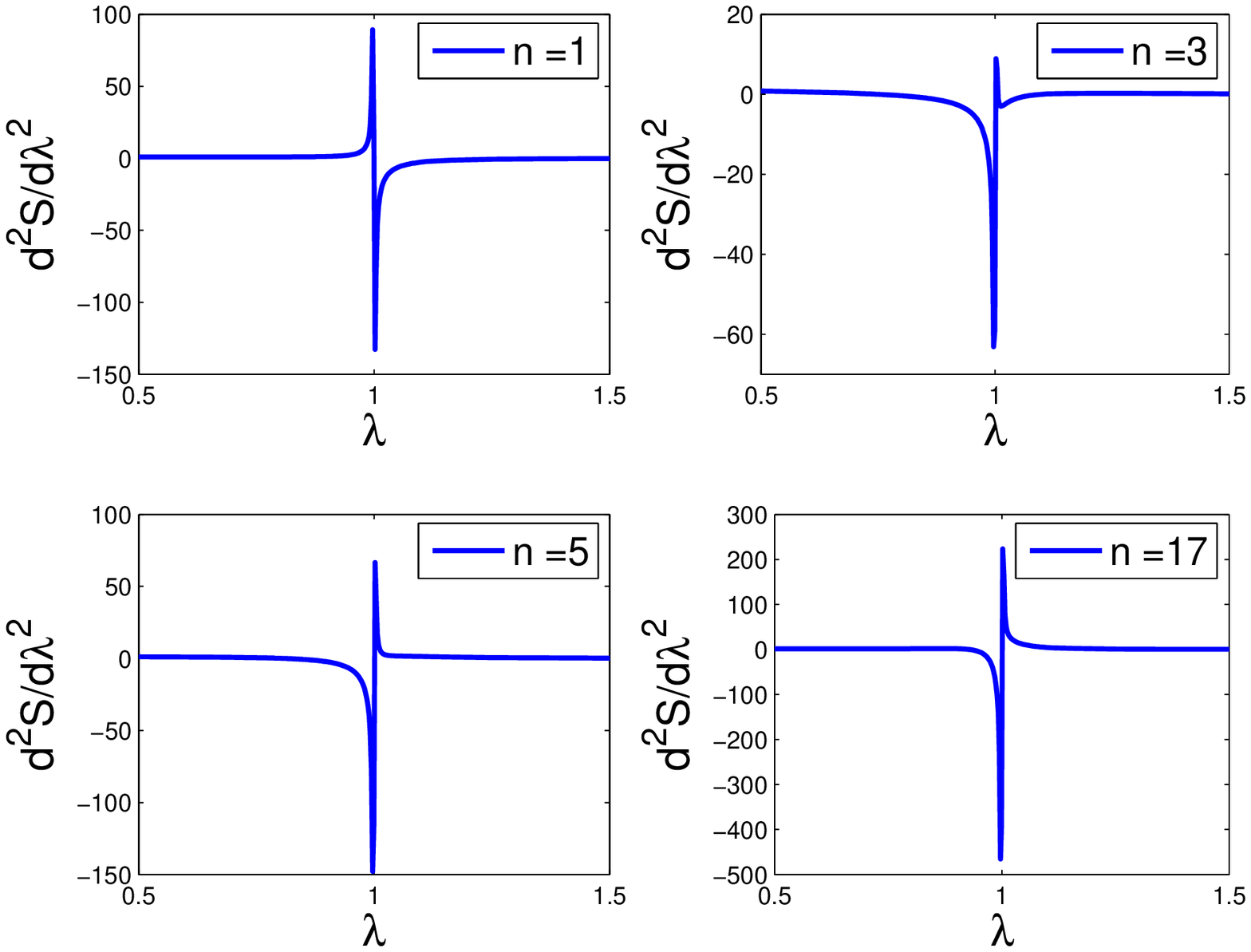} \\ (b) 
\caption{(a) The second derivative of $S(2)$ with respect to $\lambda$. This plot confirms the kink of $S(2)$ near $\lambda=1$, although the anomaly in this case is much weaker than $S(1)$. Left inset is $S$ and right inset is the first derivative. (b) The second derivative of $S(n)$ for $n=1,3,5,17$. The non-analyticity near $\lambda=1$ is evident in this plot.}\label{FIG-d2s}
\end{figure}

To further explore the connection between two-point entanglement and quantum phase transition, we plot $S(n)$ and $dS(n)/d\lambda$ as functions of $\lambda$ for different $n$ in Fig.\ref{FIG-s-ds}. As shown previously\cite{Osterloh2002,Cao2006}, the maximum of $S(n)$ for finite $n$ does not signal a possible phase transition. The sharp peak ($n=1$) or dip ($n>4$) in $dS/\lambda$ does correspond to a phase transition. More interestingly, when $n$ increases, the maximum starts to shift towards the critical point and develops a $\lambda$-like cusp at the critical point when $n$ becomes infinity. Another interesting point is that as $n$ increases, the anomaly becomes stronger and stronger as suggested by the larger and larger $|ds/d\lambda|_{\lambda=\lambda_c}$. This is another evidence that the two-point entanglement catches the important long range physics near a phase transition.  The short range two-point entanglements also exhibit anomaly near the critical point. In particular, the curve with $n=1$ displays a peak at critical point while long range ones have a dip. For $n=2,3$ the anomalies near the critical point are very weak showing crossover from strong peak of $n=1$ to strong dip for $n>4$. We also want to emphasize that $S(2)$ is non-analyticity at $\lambda=1$ although it is not as apparent as $S(1)$ in the first derivative plot. To show this, we plot the second derivative in Fig.\ref{FIG-d2s}(a), which confirms a kink near $\lambda=1$. In Fig.\ref{FIG-d2s}(b), we also plot the second derivative for different $n$. It is straightforward to see the non-analyticity  near $\lambda=1$ in this plot. With this plot, the difficulty of identifying phase transition from $n=2,3$ is circumvented. The entanglement $S(1)$ involves the correlation function $\langle \sigma^x_i\sigma^x_{i+1}\rangle$, which is coincidently the expectation value of the coupling term in the Hamiltonian\cite{Cao2006}. This implies that the peak anomaly of $dS(1)/d\lambda$ is dominated by short range interactions of the Hamiltonian. One thus might expect that its predicting power of a possible phase transition is limited and depends on the details of interactions. On the other hand, we have argued that the non-analyticity in long range two-point entanglements is dominated by a diverging length scale $\xi_E$ and is expected to be universal.

\begin{figure}
\includegraphics[width=2.5in]{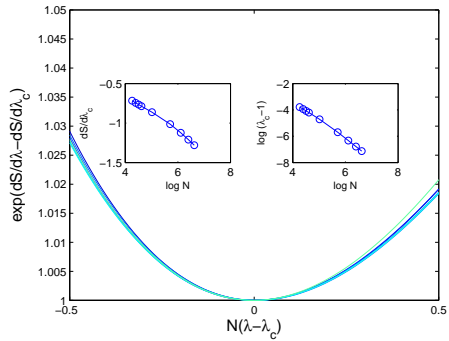}
\caption{Finite size scaling of $dS(7)/d\lambda$ with exponent $\nu=1$. The calculation was carried out for systems of difference sizes, varying from $N=61$ to $N=601$. Left inset: The first derivative at $\lambda_c(N)$ diverges logarithmically; Right inset: The position of the dip moves toward the true critical point $\lambda_c=1$ as the system size $N$ increases. Again, $\gamma=1$ in this plot. %We have verified that similar scaling law holds with critical exponent $\nu=1$ when $0<\gamma<1$ is considered.
}\label{FIG-scaling}
\end{figure}

Typical finite size scaling for $dS(n)/d\lambda$ is shown in Fig.\ref{FIG-scaling}. The data of $dS(7)/d\lambda$ collapse onto a universal line for different sample sizes $N$ and different coupling strength $\lambda$. The plot contains data from system size varying from $N=60$ up to $N=701$. The critical exponent extracted from the scaling is again $\nu=1$, in agreement with our previous discussion that the two-point entanglement $S(n)$ contains the correct physics dominated by a diverging length scale near the critical point. In the left inset of Fig.\ref{FIG-scaling}, we also plot the dip value $dS(7)/d\lambda_c$ which diverges logarithmically. The position of the dip moves toward the true critical point $\lambda_c=1$ as the system size $N$ increases. It has been shown recently that $dS(1)/d\lambda$ also bears beautiful finite size scaling with exponent $\nu=1$\cite{Cao2006}. We have also verified that similar scaling law holds with critical exponent $\nu=1$ when $0<\gamma<1$ is considered. As an example, we plot the finite size scaling of $dS(3)/d\lambda$ for $\gamma=0.5$ in Fig.\ref{FIG-scaling3}.

\begin{figure}
\includegraphics[width=2.5in]{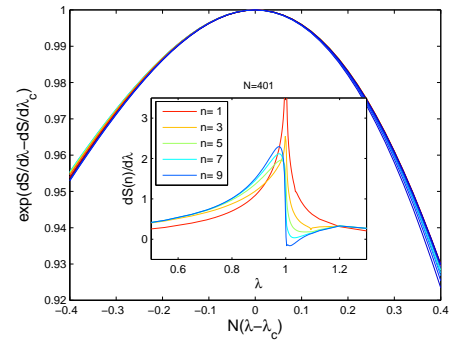}
\caption{Finite size scaling for $dS(3)/d\lambda$ with $\gamma=0.5$. The size $N$ varies from $61$ up to $1001$. Inset plots $dS(n)/d\lambda$ for different $n$ with $N=401$.}\label{FIG-scaling3}
\end{figure}

In summary, we study the two-point entanglement $S(n)$ which measures the entanglement between two separated degrees of freedom and the rest ones of the system. We introduce the entanglement length $\xi_E$ that measures how fast the entanglement saturates as the distance increases.  Two-point entanglement automatically picks out the most dominant non-analyticity of all possible two-point correlation functions. Consequently, its entanglement length $\xi_E$ agrees with the correlation length of the most dominant correlation functions. The universality and finite size scaling of entanglement are also studied in a class of exactly solvable spin model.

We would like to acknowledge useful discussions with E. Fradkin and Y. Wang.
 This material is based upon work supported by the U.S. Department of Energy, Division of Materials Sciences under Award No. DEFG02-91ER45439, through the Frederick Seitz Materials Research Laboratory at the University of Illinois at Urbana-Champaign.

{\it Note added:} After the submission of this work, we notice an alternative and very similar proposal\cite{deOliveira2006}.

%\bibliography{entanglement}

\end{document}